\documentclass[aps,prd,nofootinbib,twocolumn]{revtex4-1}
\usepackage[T1]{fontenc}
\usepackage[utf8]{inputenc}
\usepackage[english]{babel}
\usepackage{url}
\usepackage{mathrsfs}
\usepackage{amsfonts}
\usepackage{amsmath}
\usepackage{amssymb}
\usepackage{amsthm}
\usepackage{graphicx}

\usepackage[toc,title]{appendix}
\usepackage{bm}
\usepackage{booktabs}
\usepackage{braket}
\usepackage{caption}
\usepackage{enumerate}
\usepackage{enumitem}
\usepackage[Bjornstrup]{fncychap} 
\usepackage{fnpos} %
\usepackage[bottom,multiple]{footmisc} 
\usepackage{lipsum}	
\usepackage{mathtools} 
\usepackage[]{natbib}	
\usepackage{quoting}
\quotingsetup{font=small}
\usepackage{sidecap}
\usepackage{subfig}
\usepackage{textcomp}
\usepackage{tikz}
\usetikzlibrary{decorations.pathmorphing,calc}
\usepackage{titlesec}
\usepackage{tensor}
\usepackage{xfrac}

\usepackage{csquotes}
\usepackage{helvet}
\usepackage{listings}

\usepackage[bookmarks,bookmarksopen=true,colorlinks,citecolor=blue,linkcolor=blue,urlcolor=blue]{hyperref}

\makeFNbelow
\newcommand{\be}{\begin{equation}}
\newcommand{\ee}{\end{equation}}
\newcommand{\bn}{\begin{equation*}}
\newcommand{\en}{\end{equation*}}

\newcommand{\lie}{\pounds} 

\theoremstyle{plain}

\begin{document}

\rmfamily
\title{Scalar charge of black holes in Einstein-Maxwell-dilaton theory}
\date{\today}
\author{Costantino Pacilio}
\affiliation{SISSA, Via Bonomea 265, 34136 Trieste, Italy, EU}
\affiliation{INFN, Sezione di Trieste}
\begin{abstract}
We show that the monopole scalar charge of black holes in Einstein--Maxwell--dilaton theory is proportional to the electric potential at the event horizon, with a proportionality factor given by (minus) the scalar coupling constant. We also show that the scalar charge, in the weak electric charge limit, does not depend on the black hole spin. This result can be very useful to circumvent spin degeneracy issues when testing the theory against gravitational waves observations.
\end{abstract}
\maketitle
%
\section{Introduction}
\label{sec:introduction}
Compact binaries serve as ideal laboratories where to test general relativity (GR) and its modifications \cite{will_lr,berti_testing}, due to the emission of gravitational waves (GW) and, possibly, of waves from other degrees of freedom (e.g. scalar waves in scalar-tensor theories). In this respect, we can divide modified theories into (i) those which modify the propagation of the associated waves \emph{but not} the structure of the compact objects of interest; and (ii) those which modify also the structure of the compact objects. In the latter case, one says that the compact objects acquire nonstandard ``hair'', a nomenclature originated from the ``no-hair'' theorems for black holes (BHs) \cite{berti_testing,volkov} and which, for simplicity, we extend to all compact objects.

Among the possible hair, the presence of a monopole scalar charge is particularly relevant, since it triggers the emission of dipole scalar radiation. This radiative channel severely alters the energy loss balance of the binary, thus allowing in principle to extract efficient constraints from binary observations \cite{will_lr,berti_testing,yagi_1,yagi_2,berti_extreme_1}. The knowledge of the scalar monopole charge (henceforth simply ``scalar charge'') is therefore an important piece of information in the GR testing efforts. 

Recently, \cite{prabhu} showed that, in theories where the scalar charge is linearly coupled to a quadratic-curvature topological invariant or to a gauge topological invariant, it is possible to predict the scalar charge of stationary BHs without the explicit knowledge of the solution; in particular, the charge is proportional to the surface gravity of the horizon or to the magnetic potential at the horizon respectively. Using a similar argument, we show that the scalar charge can be predicted also when the scalar is coupled exponentially to the kinetic term of a gauge field: in this case, it is proportional to the electric potential energy at the horizon.

An exponential coupling of such a kind is characteristic of Einstein--Maxwell--dilaton theory (EMD) \cite{solution_1}. EMD is specified by the action
\be
\label{eq:action}
S=\int d^4x\frac{\sqrt{-g}}{16\pi}\left[R-2g^{ab}\nabla_a\Phi\nabla_b\Phi-e^{-2\eta\Phi}F_{ab}F^{ab}\right]
\ee
where $\Phi$ is a real scalar field (the dilaton) and $F_{ab}=\partial_aA_b-\partial_bA_a$ is the Maxwell tensor of the real electromagnetic (EM) field $A_a$.
EMD can be viewed as an extension of the ordinary Einstein-Maxwell theory, recovered in the limit of vanishing scalar coupling $\eta=0$. The consideration of EMD has several motivations: when $\eta=1$ it emerges in a low energy limit of string theory \cite{solution_1,solution_3}, while the case $\eta=\sqrt{3}$ corresponds to the compactification of the five-dimensional Kaluza-Klein theory \cite{solution_3,frolov_metric}. Moreover it constitutes an example of a theory where all the three bosonic degrees of freedom (scalar, vector and tensor) can be dynamical. Recently the physics of single and binary black holes in EMD has received new attention, mainly due to the perspective of observational tests offered by the emerging field of GW astronomy \cite{lehner_1,lehner_2,julie,kubzinak,io_richard}, see also \cite{ferrari_1}. Moreover, Ref.\cite{eht_dilaton} considered EMD with the addition of an axion field as an example of modified black hole, in relation to the theory-testing prospects of the Event Horizon Telescope.

A direct computation of the scalar charge in EMD BHs is partially hindered by the fact that explicit stationary solutions are known only in the static and in the slowly rotating configurations \cite{solution_1,solution_3}, with the exception of the single case $\eta=\sqrt{3}$, for which fully rotating solutions were derived in \cite{frolov_metric}. (Notably, \cite{emd_uniqueness} proved an uniqueness theorem for stationary EMD BHs under the condition $0\leq\eta^2\leq 3$.) Our result holds in complete generality and therefore encompasses also the cases in which the knowledge of the solution is still lacking.

The derivation of the scalar charge is based on a reduction of the equations of motion (EOM) to an exact rank-3 differential form. In \cite{yagi_3}, in the context of Einstein--dilaton--Gauss-Bonnet theory with a linear dilaton coupling, a similar EOM reduction was used to show that horizonless compact objects cannot support a scalar charge. We will argue that the same conclusion cannot be necessarily drawn for EMD theory.

The paper is organized as follows. In Sec.\ref{sec:formalism} we introduce the formalism of EMD theory. In Sec.\ref{sec:charge} we derive the relation between the scalar charge and the electric potential at the event horizon. Sec.\ref{sec:examples} discusses explicit examples and provides an approximate analytical expression for the scalar charge in the weak charge limit. Further details are provided in the Appendices. We adapt our notation to the mostly plus metric convention $(-+++)$ and work in units $c=G=1$.

\section{Einstein-Maxwell-dilaton theory}
\label{sec:formalism}
The EOM derived from the EMD action \eqref{eq:action} are
\begin{subequations}
\begin{align}
&S\equiv\nabla_a\nabla^a\Phi+\frac{\eta}{2}e^{-2\eta\Phi}F_{ab}F^{ab}=0\,,\\
&J_a\equiv\nabla_b\left(e^{-2\eta\Phi}F\indices{^b_a}\right)=0\,,\\
&E_{ab}\equiv G_{ab}-T^\Phi_{ab}-T^F_{ab}=0\,.
\end{align}
\end{subequations}
Here $G_{ab}$ is the Einstein tensor, while the dilaton and EM stress energy tensors are respectively
\begin{subequations}
\begin{align}
&T^\Phi_{ab}=2\nabla_a\Phi\nabla_b\Phi-(\nabla\Phi)^2g_{ab}\,,\\
&T^F_{ab}=e^{-2\eta\Phi}\left(2F_{ac}F\indices{_b^c}-\frac{1}{2}F^2g_{ab}\right)\,,
\end{align}
\end{subequations}
where we used the shorthand notations $(\nabla\Phi)^2=g^{ab}\nabla_a\Phi\nabla_b\Phi$ and $F^2=F_{ab}F^{ab}$.

For the purposes of this paper it is more convenient to express the Lagrangian and the EOM in the language of differential forms. The Lagrangian reads
\begin{subequations}
\label{lagrangian:form}
\begin{align}
& \mathbb{L}=\mathbb{L}_g+\mathbb{L}_{\Phi A}\,,\\
& 8\pi G\mathbb\,{L}_{\Phi A}=\star d\Phi\wedge d\Phi-\,e^{-2\eta\Phi}F\wedge\star F\,,
\end{align}
\end{subequations}
where $F=dA$, $\mathbb{L}_{\Phi A}$ is the dilaton-EM Lagrangian, while $\mathbb{L}_g$ is the Einstein-Hilbert Lagrangian, whose differential form we don't need explicitely in the following. Varying $\mathbb{L}_{\Phi A}$ w.r.t. $\Phi$ and $A$ we obtain the matter EOM
\begin{subequations}
\begin{align}
& \Phi : &d\star d\Phi+\eta\,e^{-2\eta\Phi}F\wedge\star F=0\,,\label{eom:1}\\
& A: &d\left[e^{-2\eta\Phi}\star F\right]=0\,.\label{eom:2}
\end{align}
\end{subequations}
Using \eqref{eom:2}, the dilaton equation \eqref{eom:1} becomes
\be
\label{eom:3}
d\left[\star d\Phi+\eta\,e^{-2\eta\Phi}A\wedge\star F\right]=0.
\ee
Now, contracting \eqref{eom:3} with a generic fixed vector field $\xi^a$ and applying Cartan's formlua $\lie_\xi=d i_\xi+i_\xi d$\footnote{Hereafter the symbol $i_\xi$ denotes the interior product of the vector field $\xi^a$ with a differential form.} on the differential form in the square brackets of \eqref{eom:3}, we obtain
\begin{multline}
\label{eom:4}
d\left[i_\xi\star d\Phi+\eta\,e^{-2\eta\Phi}i_\xi A\wedge\star F-\eta\,e^{-2\eta\Phi}A\wedge i_\xi\star F\right]\\
-\lie_\xi\left[\star d\Phi+\eta\,e^{-2\eta\Phi}A\wedge\star F\right]=0.
\end{multline}
This will be the key equation in the next section.
\section{The value of the dilaton charge}
\label{sec:charge}
Let us consider a \emph{stationary} asymptotically flat nonextremal BH solution of EMD. Then, from standard rigidity arguments (as reviewed \text{e.g.} in \cite{racz_wald_2}) the spacetime admits a timelike Killing field $t^a$, associated to time translations, and a spacelike Killing field $\psi^a$, associated with the rotational symmetry of the spacetime around the spinning axis of the BH. Then the BH horizon is a null hypersurface generated by a particular linear combination $\chi^a=t^a+\Omega_{H}\psi^a$, where $\Omega_{H}=const.$ is the angular velocity of the BH horizon. Moreover, since the BH is nonextremal, it admits a bifurcation surface $\mathcal{B}$ where $\chi^a=0$. We assume that the matter fields respect the same symmetries of the spacetime, i.e. $\lie_t A=\lie_t\Phi=0$ and similarly for $\psi^a$; then \eqref{eom:4} for $\xi^a\equiv\chi^a$ simplifies to
\be
\label{eom:5}
d\left[i_\chi\star d\Phi+\eta\,e^{-2\eta\Phi}i_\chi A\wedge\star F-\eta\,e^{-2\eta\Phi}A\wedge i_\chi\star F\right]=0.
\ee
We want to integrate \eqref{eom:5} between the BH bifurcation surface $\mathcal{B}$ and spatial infinity. In order to do this, let us specify boundary conditions for the falloff of the matter fields at infinity. By asymptotic flatness, the dilaton field will asymptote as
\be
\label{inf:scalar}
\Phi=\Phi_\infty+\frac{\Phi_1}{r}+O\left(\frac{1}{r}\right)
\ee
where $\Phi_\infty$ is a constant and $\Phi_1$ is related to the dilaton charge $Q_S$ by
\be
\label{scalar:charge}
Q_S=\int_\infty\frac{\Phi_1}{r^2}\,\epsilon_2
\ee
with $\epsilon_2$ the area element at infinity. We choose the EM gauge such that the vector field scales as
\be
\label{a:falloff}
A_a\,dx^a=\frac{\bar{A}_a}{r}dx^a+O\left(\frac{1}{r}\right)\,.
\ee
Then the electric potential is given by $V_E=i_\chi A|_\mathcal{B}$ (notice that, despite $\chi^a|_\mathcal{B}=0$, $V_E$ can be finite if $A$ diverges on $\mathcal{B}$). It can be easily proven that $\lie_\chi A=0$ implies that $V_E$ is constant over the event horizon and that the projection of $i_\chi\star F$ on $\mathcal{B}$ vanishes (see Appendix\,\ref{sec:horizon}).

We are ready to integrate \eqref{eom:5}. Using Gauss' theorem, the integration reduces to two surface contributions respectively at infinity and at $\mathcal{B}$. The contribution at infinity gives
\begin{multline}
\label{eq:integral}
\int_\infty\left(i_\chi\star d\Phi+\eta\,e^{-2\eta\Phi}i_\chi A\wedge\star F-\eta\,e^{-2\eta\Phi}A\wedge i_\chi\star F\right)\\
=\int_\infty i_\chi\star d\Phi=4\pi\,Q_S\,.
\end{multline} 
In the first step we used the asymptotic falloff \eqref{a:falloff} to cancel the terms involving the EM field, while in the second step we used the fact that $\psi^a$ is tangential to the integration surface. The contribution at the bifurcation surface reads
\begin{multline}
\int_\mathcal{B}\left(i_\chi\star d\Phi+\eta\,e^{-2\eta\Phi}i_\chi A\wedge\star F-\eta\,e^{-2\eta\Phi}A\wedge\star i_\chi F\right)\\
=\eta\int_\mathcal{B}e^{-2\eta\Phi}i_\chi A\wedge\star F=-4\pi\eta\,Q_EV_E
\end{multline}
where we used $\chi|_\mathcal{B}=0$, $i_\chi\star F|_\mathcal{B}=0$ and the definition of the electric charge
\be
Q_E=-\frac{1}{4\pi}\int_\mathcal{B}e^{-2\eta\Phi}\star F\,.
\ee
Therefore, putting the two contributions together, we arrive at the main result of this paper\footnote{The same result, but restricted to the static case, has been recently obtained in \cite[Eq.(4.19)]{nozawa}.}
\be
\label{scalar:charge}
\boxed{Q_S=-\eta\,W_E}
\ee
where $W_E=Q_EV_E$ is the electric potential energy of the EM field. $\square$

Let us emphasize that, if we assume the existence of a bifurcation surface, the result does not depend explicitely on the Einstein EOM (see Appendix\,\ref{sec:horizon}); therefore the restriction that $\mathbb{L}_g$ in \eqref{lagrangian:form} is the Einstein-Hilbert Lagrangian can also be relaxed. 

Notice that $Q_S$ does not depend on the value $\Phi_\infty$ of the dilaton field at spatial infinity. It is then clear from \eqref{scalar:charge} that the scalar charge is a secondary hair,\footnote{Following \cite{herdeiro_radu}, we call ``secondary hair''  those which are not independent, but depend on the other global charges, in this case the mass and the electric charge.} as it could have been guessed from the fact that no Gauss law is associated to the dilaton field. However, as observed in \cite{emd_1law_1,emd_1law_2}, it enters the dynamical first law by sourcing variations of the asymptotic value $\Phi_\infty$ if the latter is allowed to vary,
\be
\delta M=\frac{\kappa}{8\pi}\delta A_H+\Omega_H\delta J+V_E\delta Q_E-Q_S\delta\Phi_\infty\,.
\ee
\paragraph*{Horizonless compact objects.}
If the BH is replaced by an horizonless compact object, e.g. a neutron star, one might be tempted to integrate \eqref{eom:5} over a compact hypersurface with a single boundary at spatial infinity, thus showing that the scalar charge of the object is necessarily zero. However, this reasoning would be fallacious. Indeed, for an electrically charged BH in EMD to originate from a collapse, the collapsing material must already support the electric charge within its degrees of freedom. Therefore $\mathbb{L}_{\Phi A}$ must be supplied by an interaction Lagrangian between ordinary matter and the vector field $A_a$, \emph{even if $A_a$ is not the photon field of the standard model.} In a nonvacuum spacetime such a coupling will induce an additional electric current term in the EOM \eqref{eom:2}. In turn, this will generate a volume contribution to the integral \eqref{eq:integral}, thus spoiling the above reasoning.
\section{Explicit examples}
\label{sec:examples}
In this section we provide explicit examples of the identity \eqref{scalar:charge}. We will refer to the analytic BH solutions already existing in the literature \cite{solution_1,solution_2,solution_3,frolov_metric,larsen_metric}. Static and slowly rotating BH solutions are known for all the values of the dilaton coupling $\eta$ \cite{solution_1,solution_2,solution_3}, while fully rotating solutions are known explicitely only for $\eta=\sqrt{3}$ (and trivially also for $\eta=0$, in which the BH is described by the well known Kerr-Newman solution) \cite{solution_3,frolov_metric,larsen_metric}. The value of the scalar charge for a single rotating BH in EMD was estimated numerically in \cite{lehner_2}, finding that the charge decreases as the spin increases; however, we anticipate that our analytic formulae \eqref{scalar:charge} and \eqref{scalar:charge:3} disagree with such a conclusion in the limit of weakly charged BHs, a limit which is consistent with the approximations of \cite{lehner_2}.

Static spherically symmetric BH solutions in EMD are described by the line element \cite{solution_1, solution_2,solution_3}
\begin{subequations}
\label{eq:line:1}
\begin{align}
&ds^2=-F(r)dt^2+\frac{dr^2}{F^(r)}+r^2G(r)\left(d\theta^2+\sin^2\theta d\varphi^2\right)\,,\\
&F(r)=\left(1-\frac{R_+}{r}\right)\left(1-\frac{R_-}{r}\right)^{(1-\eta^2)/(1+\eta^2)}\,,\\
&G(r)=\left(1-\frac{R_-}{r}\right)^{2\eta^2/(1+\eta^2)}\,.
\end{align}
\end{subequations}
Here $R_+$ and $R_-$ are, respectively, the radii of the outer and inner horizons
\begin{subequations}
\label{eq:rpm:1}
\begin{align}
&R_+=M\left(1+\sqrt{1-(1-\eta^2)\,\sigma^2}\right)\,,\\
&R_-=M\left(\frac{1+\eta^2}{1-\eta^2}\right)\left(1-\sqrt{1-(1-\eta^2)\,\sigma^2}\right)\,.
\end{align}
\end{subequations}
$M$ is the mass and $\sigma$ is the electric charge-to-mass ratio, $\sigma=Q_E/M$. The scalar and the vector fields are given by
\begin{subequations}
\label{eq:matter:1}
\begin{align}
&\Phi=\frac{\eta}{1+\eta^2}\log\left(1-\frac{R_-}{r}\right)\,,\label{phi:1}\\
&A_a\,dx^a=\frac{M\sigma}{r}\,dt\,.
\end{align}
\end{subequations}
Notice that the action is invariant under the reparametrization
\be
\label{eq:rep}
A_a\to e^{-\eta\Phi_0}A_a\qquad \Phi\to\Phi -\Phi_0
\ee
where $\Phi_0$ is a constant. For simplicity, we choose $\Phi_0$ such that $\Phi\to0$ at spatial infinity.

The generator of the BH horizon is the Killing field $t^a$ and therefore the EM potential is simply $V_E=Q_E/R_+$. If we expand \eqref{phi:1} in series of $1/r$, we find that the leading term is
\be
\lim_{r\to\infty}\Phi=-\frac{\eta}{1+\eta^2}\frac{R_-}{r}+O\left(\frac{1}{r}\right).
\ee
Therefore the scalar charge is
\begin{multline}
\label{scalar:charge:2}
Q_S\equiv\Phi_1=-\frac{\eta\,R_-}{1+\eta^2}=-\frac{\eta M\left(1-\sqrt{1-(1-\eta^2)\sigma^2}\right)}{1-\eta^2}\\
=-\frac{\eta M\,\sigma^2}{1+\sqrt{1-(1-\eta^2)\sigma^2}}=-\eta\frac{Q_E^2}{R_+}
\end{multline}
in agreement with Eq.\,\eqref{scalar:charge}. Before going on, notice that the first corrections to metric relative to Schwarzschild appear at $O(\sigma^2)$. Therefore we \emph{define} the weak charge limit as the expansion of the fields at second order in the electric charge. At this order $Q_S=-\eta\,Q_E^2/2M$.

The weak charge limit proves particularly useful when we switch to rotating configurations, because it allows to overcome the lack of explicit solutions. Indeed, if we assume a weak charge $Q_E$, and since the static dilaton charge is already quadratic in $Q_E$, we can neglect higher order corrections coming from the fully rotating metric, thereby reducing the problem to the usual Kerr metric.

More specifically, the argument goes as follows. The electric potential energy in the rotating configuration will have the form
\be
W_E=\frac{Q_E^2}{2M}\times f(M,a,Q_E,\eta)
\ee
where $f(M,a,Q_E,\eta)$ is a factor accounting for corrections due to the nonvanishing spin $a$, such that $f(M,0,Q_E,\eta)=1$. Since we are interested in the weak charge limit, we can neglect the corrections due to $Q_E$; however, since $\eta$ is always accompained by factors of $Q_E$,\footnote{The reader can convince herself that it must be so because, if the electric charge vanishes, all the terms in EOM containg $\eta$ vanish as well.} we can also neglect the dependence on $\eta$. Therefore we reduce to $f(M,a)$ as given by the Kerr metric, i.e.
\be
\label{scalar:charge:3}
W_E=\frac{Q_E^2\,R_+}{R_+^2+a^2}=\frac{Q_E^2}{2M}+O\left(Q_E^2\right)\,.
\ee
We have thus obtained the important result that, in the weak charge limit, the dilaton charge \emph{does not} depend on the spin. As we anticipated, this conclusion contrasts with Eq.(39) of \cite{lehner_2}, which was obtained in the particular case $\eta=1$. This discrepancy may be due to the propagation of numerical approximation errors in the initial conditions, see the first paragraph of Sec.IV.A in \cite{lehner_2}.

The nondependence on the spin in \eqref{scalar:charge:3} gives a great advantage in the context of theory testing. Indeed, GW events so far constrained the BH spins only loosely, thus jeopardizing the effectiveness of the tests based on the emission of scalar dipole radiation \cite{yagi_2}. The fact that \eqref{scalar:charge:3} is independent of the spin suggests that less degenerate theoretical bounds can be obtained.

Curiously, the nondependence on the spin persists at all orders in $Q_E$ in the fully rotating Kaluza-Klein solution (see Appendix\,\ref{sec:sqrt3}). Given that this is the only nontrivial value of the dilaton coupling for which fully rotating BHs are known, we do not find it enough to speculate that $Q_S$ is spin independent for all values of $\eta$.
\section{Discussion}
In this paper we have shown that the monopole scalar hair of stationary BHs in Einstein--Maxwell--dilaton theory is related to the electric potential at the event horizon. Although our procedure is similar to the one in \cite{yagi_3,prabhu}, we cannot draw the additional conclusion that horizonless compact objects cannot support a monopole scalar charge. However, only in the case of BHs we are able to express the scalar charge in a simple form in terms of the mass and the electric charge, Eq. \eqref{scalar:charge}.

The main assumption behind our result is that all the dynamical fields are Lie dragged along the Killing field $\chi^a$ generating the event horizon: this allows the second line of \eqref{eom:4} to vanish. It might be possible that the relation \eqref{scalar:charge} still holds for nonstationary configurations, finely tuned in such a way that the expression in square brackets in the second line of \eqref{eom:4} is still Lie dragged along $\chi^a$. We are not aware of a solution of the EMD EOM with this property (see however \cite{rocha} for a discussion of nonstationary EMD BH solutions).

It must be stressed that, although the vector field in EMD does not necessarily coincide with the photon field of the standard model, the guiding principle of small deviations from GR suggests that the electric charge, if nonvanishing, is small. We provided an operative meaning of the ``weak electric charge limit'' and showed that the scalar charge in this limit can be computed in a closed form, even for fully rotating configurations. Moreover, in this limit, the scalar charge does not depend on the spin.

These results are certainly useful to test EMD in the inspiral phase of binary events \cite{berti_extreme_1}, where the spin degeneracy is one of the main issues in deriving sensible constraints. A complementary analysis would be the study of the ringdown oscillation modes, which is presently limited to the $\eta=1$ case \cite{ferrari_1}. This will be the subject of a work in preparation \cite{io_richard}.
\begin{acknowledgements}
The author thanks S. Liberati and R. Carballo-Rubio for fruitful discussions about an earlier version of the manuscript.
\end{acknowledgements}
\begin{appendices}
\section{The Maxwell field at the event horizon}
\label{sec:horizon}
We summarize the properties of the EM field on the event horizon, following the same arguments presented in \cite{carter:book}, and in \cite{isolated_1,isolated_2} in the context of isolated horizons. Since the horizon is generated by a Killing field $\chi^a$, its expansion and shear vanish identically. Then the Raychauduri equation implies $R_{ab}\chi^a\chi^b=0$ at the horizon. Using the gravitational EOM and the fact that $\lie_\chi\Phi=\chi^a\nabla_a\Phi=0$, we find that the vector $\chi^aF_{ab}$ must be null on the horizon. Since $F_{ab}\chi^a\chi^b=0$, this means that the pullback of $i_\chi F$ on the horizon vanishes. The same argument can be repeated for $i_\chi\star F$, once one notices that the Maxwell stress energy tensor can be rewritten as
\be
2F_{ac}F\indices{_b^c}-\frac{1}{2}F^2g_{ab}=2(\star F)_{ac}(\star F)\indices{_b^c}-\frac{1}{2}(\star F)^2g_{ab}
\ee
where $\star F_{ab}=\epsilon_{abcd}F^{cd}/2$. Now,
\be
\label{v:const}
0=\lie_\chi A=d(i_\chi A)+i_\chi F
\ee
and therefore, using the fact that the pullback of $i_\chi F$ on the horizon vanishes, we deduce that $V_E=i_\chi A=\text{const.}$ everywhere on the horizon.

The above arguments do not make reference to the bifurcation surface $\mathcal{B}$, but uses explicitely the Einstein equations. Alternatively \cite{prabhu_2}, if we assume that a bifurcation surface exists, then $i_\chi F=0=i_\chi\star F$ on $\mathcal{B}$, from which it follows through \eqref{v:const} that $V_E=\text{const.}$ on $\mathcal{B}$. Finally, using $\lie_\chi A=0\implies\lie_\chi(i_\chi A)=0$, the constancy of $V_E$ on the whole horizon is established.
\section{Rotating $\eta=\sqrt{3}$ black holes}
\label{sec:sqrt3}
Fully rotating EMD BHs are known for $\eta=\sqrt{3}$ \cite{solution_3,frolov_metric,larsen_metric}. The line element is (we follow the notation of \cite{solution_3})
\begin{multline}
ds^2=-\frac{1-Z}{B}dt^2-2a\frac{Z\,\sin^2\theta}{B\sqrt{1-w^2}}dtd\varphi+\\
\left[B(r^2+a^2)+a^2\sin^2\theta\frac{Z}{B}\right]\sin^2\theta d\varphi^2+B\frac{\Sigma}{\Delta}dr^2+B\Sigma d\theta^2
\end{multline}
where
\begin{subequations}
\begin{align}
&\Sigma=r^2+a^2\cos^2\theta\,,\qquad\Delta=r^2+a^2-2m\,r\,,\\
&Z=\frac{2m\,r}{\Sigma}\,,\qquad B=\left(1+\frac{w^2\,Z}{1-w^2}\right)^{1/2}\,.
\end{align}
\end{subequations}
The vector field and the dilaton field are given by, respectively,
\begin{subequations}
\begin{align}
&A_a\,dx^a=\frac{w\,Z}{2(1-w^2)B^2}\left[dt-a\sqrt{1-w^2}\sin^2\theta\,d\varphi \right]\,,\\
&\Phi=-\frac{\sqrt{3}}{2}\ln B\,.\label{phi:2}
\end{align}
\end{subequations}
The mass $M$, the electric charge $Q_E$ and the angular momentum $J$ are given by
\begin{subequations}
\begin{align}
&M=m\left(1+\frac{w^2}{2(1-w^2)}\right)\,,\\
&Q_E=\frac{m\,w}{1-w^2}\,,\\
&J=\frac{m\,a}{\sqrt{1-w^2}}\,.
\end{align}
\end{subequations}
The timelike and rotational Killing fields are $t^a=(\partial/\partial t)^a$ and $\psi^a=(\partial/\partial \varphi)^a$, while the Killing generator of the horizon is $\chi^a=t^a+\Omega_H\psi^a$, with
\be
\label{eq:omega}
\Omega_H=-\left.\frac{g_{t\varphi}}{g_{\varphi\varphi}}\right|_{\theta=\pi/2}=\frac{a\,\left(1-w^2\right)^{3/2}}{r_+^2+a^2}
\ee
and $r_+$ is the outermost solution of $\Delta=0$. By expanding \eqref{phi:2} in powers of $1/r$ we find that the scalar charge is
\begin{multline}
\label{scalar:charge:4}
Q_S=-\frac{\sqrt{3}}{2}Q_E\,w=-\frac{\sqrt{3}}{2}\left(\sqrt{M^2+2Q_E^2}-M\right)\\
=-\sqrt{3}\,\frac{Q_E^2}{2M}+O\left(Q_E^2\right)\,.
\end{multline}
Expression \eqref{scalar:charge:4} is independent of $a$, it coincides with \eqref{scalar:charge:2} for $\eta=\sqrt{3}$ and, using \eqref{eq:omega}, the reader can easily verify that it respects the relation \eqref{scalar:charge}.
\end{appendices}

\end{document}